\documentstyle[prb,floats,aps,amssymb,eqsecnum,preprint,epsf,rotate]{revtex}
\begin{document}

\draft

\title{Study of the one-dimensional off lattice hot monomer reaction
model.}
\author{Daniel H. Linares}
\address{Departamento de F\'{\i}sica and Centro Latinoamericano de Estudios 
Ilya Prigogine, Universidad Nacional de San Luis, Chacabuco 917, C. C. 136, 
5700 San Luis, Argentina}  
\author{Ezequiel V. Albano and  Roberto A. Monetti}
\address{Instituto de Investigaciones Fisicoqu\'{\i}micas Te\'oricas y 
Aplicadas (INIFTA), UNLP, CONICET, CIC (Bs. As.), C. C. 16 Suc. 4,
1900 La Plata, Argentina} 
\date{\today}

\maketitle

\begin{abstract}
Hot monomers are particles having a transient  mobility (a ballistic flight) 
prior to being definitely absorbed on a surface. After arriving at
a surface, the excess energy coming from the kinetic energy in the gas
phase is dissipated through degrees of freedom parallel to the surface
plane. In this paper we study the hot monomer-monomer
adsorption-reaction process on a continuum (off-lattice) one-dimensional space
by means of Monte Carlo simulations. 
The system exhibits a second-order irreversible phase transition between
a reactive and saturated (absorbing) phases which belong to the
directed percolation (DP) universality class. This result is interpreted by
means of a coarse grained Langevin description which allows as to extend
the DP conjecture to transitions occurring in continuous media.
 
\end{abstract}

\pagebreak
\narrowtext

\narrowtext

\section{Introduction}
Interacting particle systems are relevant to wide-ranging phenomena in 
physics, chemistry, biophysics, ecology, etc. The concept of
'particles' is used in a broad sense, that is 
'particles' can be atoms, molecules, spins, individuals, etc, and
whilst attention is
drawn to the interactions among particles no attempt is made in order to
achieve a detailed description (e. g. quantum mechanical) of the particle 
itself. Therefore, due to interactions, the occurrence of complex behavior, 
such as phase transitions, self-organization, chaos, bistability, etc, may 
be observed \cite{Ligget}. 

Within this context, an active field of research is the study of
far-from-equilibrium reaction systems \cite{ZGB,Eze1}. Irreversible
phase 
transitions (IPT) between active regimes (where reactions are
sustained) and absorbing states (where 
reactions are no longer possible) have been reported in a great variety of 
models such as the Ziff, Gulari, and Barshad (ZGB) model for the
catalytic oxidation of CO \cite{ZGB}, the 
dimer-dimer model \cite{Eze2}, the contact process \cite{Jens1},
forest-fire models \cite{Eze3}, etc (for a recent review see
e. g. \cite{Eze1}). According to the Janssen-Grassberger conjecture
\cite{Janss,Grass1}, irreversible reaction systems that exhibit a
phase transition to a single absorbing state characterized by a scalar
order parameter belong to the Directed Percolation (DP) universality
class. This conjecture, stated a long time ago for unique
absorbing states, has 
been further generalized for the cases where such states are
non unique 
\cite{Jens2,Eze4}. A special case corresponds to non-equilibrium systems
where, provided an IPT exists, there is in addition a local or global
conservation of particles modulo two, such as the branching
and annihilating random walks with an even number of offsprings
\cite{BARWO,Jens4}. In these cases a new universality class emerges, commonly called 
parity conserving (PC) class, which is due
to the existence of two statistically equivalent absorbing states at
the critical point \cite{Kolea}. However, global conservation of
particles of modulo two
may also lead to exponents in the PC class only when local spontaneous annihilation
($1X \rightarrow 0$) is highly inhibited. Then, at a coarse grained
level, the relevant surviving processes are those conserving
parity. In other words, parity conservation can 
be restored at a coarse grained level \cite{Haye,Roberto}. A nice example where global
parity conservation still leads to DP exponents is given by Inui et. al. \cite{Ponja}. 
It is clear in this case that spontaneous annihilation must be taken
into account.

IPT are studied largely by means of Monte Carlo simulations
and mean-field approaches. Recent developments of field-theoretic
renormalization group techniques have provided a new theoretical framework
where non-equilibrium phase transitions can be studied \cite{Uwe}. These
techniques are able to identify the relevant processes in a given
universality class although the quantitative predictions are still
poor. 

So far, most of the simulations have been performed using
\underline{discrete} lattices where each particle fills a single site
on the lattice and neighboring particles may react with a certain
probability. In contrast, our knowledge on the behavior of
irreversible reaction systems in continuous media is rather poor. In
order to stress some dramatic differences that may arise for a reaction system
when it is simulated off lattice, let us consider the simplest case of
the $B + B \rightarrow 0$ irreversible reaction which proceeds according
to the Langmuir-Hinshelwood mechanism.
\begin{eqnarray}
 & B(g) + S \rightarrow B(a) \\ \nonumber  
 & B(g) + S \rightarrow B(a)  \\   
 & B(a) + B(a) \rightarrow 0 + 2 S \nonumber  
\end{eqnarray}
where $g$ and $a$ refer to the gas and adsorbed phases, respectively,
while $S$ represents a site on the surface. At first, we assume that
$B$-species adsorbed on nearest neighbor sites react with unitary
probability ($P_r = 1$). If we used a discrete lattice, reactions
would be sustained indefinitely, i. e. the system could not irreversibly
evolve into an absorbing state. However, considering a continuous
media, the random adsorption of $B$ particles of finite size $\sigma$
causes the formation of several interparticle gaps of size smaller
than $\sigma$. So, in the infinite time limit ($t \rightarrow \infty$) the
sample becomes imperfectly covered by $B$-species separated by small
interparticle gaps. Reaction is no longer possible and the system
becomes irreversible trapped into an absorbing state (infinitely
degenerated). The maximum jamming coverage attained in one dimension
is $\Theta_j \approx 0.74759$, which corresponds to the so called car
parking problem \cite{Jim,car}. 

In this paper we show that by introducing the adsorption of species
with transient mobility in a continuous one-dimensional medium, it is
possible to open a window where reactions are sustained. However, by 
tuning the external parameter which controls the transient mobility
of the particles it is possible to irreversible drive the system into
an absorbing state. 

It should be mentioned that the study of reactions
of atoms in the gas phase possessing thermal energy with adsorbed
atomic CO species on metal and 
semiconductor surfaces is a topic of current interest. In contrast to
thermally activated reactions among adsorbed species, i. e., the so
called Langmuir-Hinshelwood mechanism, these kind of reactions take
place under far-from-equilibrium conditions. Consequently, the
determination of the underlying mechanism as well as the understanding
of the dynamic behavior is challenging work. Within this context,
very recently Kim {\it et al} \cite{Kim} have reported experimental studies
of the reaction of hot $H$-atoms with adsorbed $D$-atoms (for further
experimental works see references in \cite{Kim}).

It should be noted that from the theoretical point of view a number of 
related models for random sequential adsorption with diffusion
\cite{priv,eis} and desorption \cite{stinc} have also been proposed and studied
(for a review see also \cite{Jim}). However, interest in
such studies is addressed to the asymptotic approach to the jammed state.
In contrast, in this paper our interest is focused on the irreversible
critical behaviour of a reactive system.

So this work
is devoted to the characterization of 
such IPT in continuous media and it is organized as follows: section
2 gives the description of the model and the simulation technique. In
section 3 we discuss the results while conclusions and remarks are
presented in section 4.

\section{The Model and the Monte Carlo simulation method}

In this paper, we study a 1D off-lattice
adsorption-reaction model in which particles of size $\sigma$ undergo a 
ballistic flight just after depositing on the substrate.
The system evolves in time under the following local rules. (i) A position $x$
is randomly selected on the substrate. If the interval $[x - \sigma/2, x + \sigma/2]$ is 
empty, then the adsorption trial is successful, otherwise it is rejected. So, it is
clear that double occupancy of positions is forbidden. (ii) Right after a successful
adsorption trial, a random direction is selected (left or right). Then, the particle 
undergoes a ballistic flight in the previously selected direction up to a distance $R$ 
from the adsorption position $x$, provided that no other already deposited particle is found 
along the way. (iii) If during the flight one particle hits another
previously adsorbed particle which is 
already at rest on the substrate, the following alternatives can occur: (1) the annihilation 
($B+B \rightarrow 0$) occurs with probability $P_r$. Then, both particles react and leave 
the system. (2) Particles do not react (with probability $(1- P_r)$), and the flying particle 
is frozen at the collision point.

The ballistic flight mimics 'hot monomer' adsorption, allowing the incoming particle to 
transform its energy into degrees of freedom parallel to the
substratum. The length of the flight $R$ is finite in order to account for
frictional dissipation. The model has two externally tunable
parameters, namely $R$ and $P_r$. For $P_r = 0$ one recovers the 'hot
monomer' random sequential adsorption model \cite{Daniel} while for
$R = 0$ and $P_r = 0$ one has the 1d car parking problem \cite{car}.

In order to simulate a continuous medium on a digital computer, one
actually considers a discrete lattice. However, each site of size
$\sigma$ is subdivided into $2^{64}$ different adsorption
positions. This high degree of discretization has provided excellent
results when compared with the exact analytic solution of a related
problem \cite{Daniel}. 

Preliminary results show that the system can undergo continuous IPT
between a stationary 
reactive state and an absorbing state without reactions when varying
the parameters. This can easily be tested by considering the case
$P_r = 1$ and $R = 0$ ($R > 1$) which gives an absorbing (reactive)
state, respectively. 
It should be pointed out that continuous IPT are dominated by
fluctuations. Consequently, in a finite
system and close to the critical point, the stationary state of the reactive
phase can irreversibly evolve into the saturated state (absorbing state). Due
to this circumstance, the precise determination of both critical points and
critical exponents is rather difficult. However, this shortcoming can be 
avoided by performing an epidemic analysis. 
For this purpose one starts, at $t=0$, with a a configuration close to
one of the absorbing states. It is clear that different absorbing states will normally differ
in the density of monomers. It should be pointed out that the dynamical critical behavior
of systems with infinitely many absorbing configurations is expected to depend upon the
initial density of inactive particles \cite{Jens2,Mendes}. However, static critical behavior appears
to be independent of it. 
From the set of possible initial densities, 
a value $\rho_n$ is particularly important, namely the stationary density of inactive particles
which results after the natural evolution of the system in the subcritical region has finished.
The value $\rho_n$ is relevant, since only for this value the natural dynamical critical 
behavior emerges.  Preliminary simulation results show that $\rho_n$ depends on the parameter 
$P_r$, but their values have not been included in this work for the sake of space.
The dependence of the critical behavior on the set of initial densities is 
the subject of an ongoing investigation.

Consequently, the initial state for the epidemic analysis is 
generated by the evolution of the system very close to the critical
point until poisoning is achieved. After generating this stationary
poisoned state, we remove one or two particles 
from the middle of the system in order to create a small active
area where adsorption is now possible. It should be noted that an
empty area is considered to be active 
if it is longer than or equal to $\sigma$. Then, the time evolution of
the system is analyzed by measuring the following properties: (i) the
average amount of active area at time $t$, $A(t)$; (ii) the survival
probability of the active area at time $t$, $P_s (t)$; and (iii) the average
distance over which the active area have spread at time $t$,
$D(t)$. Finite size effects are absent because the system is taken
large enough to avoid the presence of active area at the boundaries. 
For this purpose a sample of $10^4 \sigma$ is enough. Averages are taken over 
$10^5$ to $10^6$ different samples. Near the critical point, the
amount of active area is often very small. Then, we improve the
efficiency of the algorithm by keeping a list of the positions where
there is active area. Time is incremented by $1/a(t)$, where $a(t)$
is the amount of active area at time $t$. The time evolution of the active
area is monitored up to $t =10^5$. At criticality, the following
scaling behavior holds: 
\begin{equation}
\label{Eta}
A(t) \propto t^{\eta},
\end{equation}
\begin{equation}
\label{Delta}
P_s(t) \propto t^{-\delta},
\end{equation}
and
\begin{equation}
\label{Zeta}
D(t) \propto t^{z/2}
\end{equation}
where $\delta $, $\eta $ and $z$ are \underline{dynamic} exponents. 

\section{Results and discussion}
Preliminary simulations show that for $P_r = 1$ it is possible to
achieve a stationary reactive state in the large $R$ limit (i. e. for
$R \ge 2$) while in the opposite limit ($R \le 1.5$) the system
becomes irreversible saturated by $B$-species. In order to obtain a
quantitative description of the IPT we have performed epidemic studies
around the critical edge. Figure 1 (a - c) shows log-log plots of $A$, $P_s$
and $D$ versus the time $t$, obtained for different parameter values. The 
three plots exhibit a power law behavior which is the signature of a
critical state. Using smaller (greater) $R$-values we observe slight
upwards (downwards) deviations in three plots which indicate
supercritical (subcritical) behavior (these results are not shown for
the sake of clarity). Then, the critical exponents obtained by
regressions are:
\begin{equation}
\label{expon}
\eta =  0.308 \pm 0.004 \; \; \; \delta = 0.165 \pm 0.003 \; \;
\; z/2 = 0.625 \pm 0.002 
\end{equation}
These values are in excellent agreement with the exponents
corresponding to the DP universality class in $1+1$
dimensions \cite{Janss,Grass1}. Recently, extended series expansions calculations \cite{series}
have provided very accurate values for the DP critical exponents, namely
\begin{equation}
\label{dpexpon}
\eta = 0.31368(4) \; \; \; \delta = 0.15947(3) \; \; \; z/2 = 0.63261(2)
\end{equation}
Therefore we conclude that the studied adsorption-reaction model on a
\underline{continuous medium} belongs to the DP universality class
like many other systems already studied on \underline{discrete
lattices}. It should be noticed that the present model has infinitely
many absorbing states, so as in the case of the dimer-dimer model
\cite{Eze2} the DP conjecture holds for non-unique absorbing states
\cite{Janss} at least as long as the absorbing states can be solely 
characterized by the vanishing of a single scalar order parameter. 

We have also studied the case of imperfect reaction, i. e. $P_r <
1$. Figure 2 shows a plot of the phase diagram. The phase boundary
curve was determined by means of an epidemic analysis, as is shown
in figure 1. The obtained critical exponents are
\begin{equation}
\label{imexpon}
\eta = 0.312 \pm 0.004 \; \; \; \delta = 0.157 \pm 0.003\; \; \; z/2 =
0.631 \pm 0.001 
\end{equation}
Once again, these exponents are in good agreement with those corresponding
to DP. Scanning the whole critical curve we obtain second order IPT's
that belong to the DP universality class. However, the special case
$R \rightarrow \infty$ merits further discussion. For $P_r = 1$ ($P_r =
0$) the system evolves towards a reactive (absorbing) state,
respectively. Then, a transition is expected at some intermediate
$P_r$ value. In this case, the time evolution of the active area can
be described by means of a mean field equation. In fact, the active
area $A(t)$ will grow (decrease) proportionally to $A(t) P_r$ ($A(t) (1 - P_r)$),
respectively; so
\begin{equation}
\label{mfield1}
\frac{d A}{dt} = A(t) P_r - A(t) (1 - P_r)
\end{equation}
which leads to
\begin{equation}
\label{mfield3}
A(t) = A_{0} \; e^{(2 P_r - 1)t} 
\end{equation}
Therefore, $P_r = 1/2$ is a critical value such as for $P_r > 1/2$
($P_r < 1/2$) the active area will increase (decrease) exponentially
in time, while just at criticality $A(t)$ will remain constant ($A(t)
= A_0$), which is consistent with a mean field exponent $\eta_{MF} =
0$. The predicted behavior is confirmed by means of simulations as it
is shown in figure 3. By means of linear regressions the following
exponents are obtained for $P_{r} = \frac{1}{2}$:
\begin{equation}
\label{imexp}
\eta \approx 0.0  \; \; \; \delta \approx 1.0 \; \; \; z/2 \approx 1.0
\end{equation}
Then, our mean field estimate for $\eta$ is in good agreement with the
simulation results. Regrettably, we were unable to derive the mean
field values for the remaining exponents.

We conclude that the particular point $P_r =
1/2$, $R \rightarrow \infty$ is a first order point (see figure 2)
which is not in the DP class which characterizes the whole critical curve. 

In the following, we give theoretical arguments by means of a coarse-grained 
Langevin description that support the result concerning the
universality class of the model. First, note that the normalized variables
needed to characterize the configurations of the system are the amount
of active area $a(x,t)$, the number of monomers in the system $n(x,t)$ and
the amount of inactive area $v(x,t)$. These variables are not
independent since we have the constrain $a(x,t) + n(x,t) + v(x,t) = 1$. It
is clear from the above discussion that the time evolution of the
system ends when $a(x,t) = 0$. Since $a(x) \rightarrow 0$ at
criticality, this quantity can be chosen as the order parameter of
the system. Then, we will try to  describe the time
evolution of the system near the critical point by means of two
coupled Langevin equations, for instance, one for $a(x,t)$ and the
other for $n(x,t)$. Due to the nature of the absorbing configurations,
each term of these equations must vanish when $a(x,t) \rightarrow
0$. \\ Let us consider the microscopic processes which are 
relevant to characterize the critical behavior of the system. First of
all, both diffusion of $a(x)$ and $n(x)$ can be interpreted as
successive adsorption-reaction processes. Within a one site
description, both $n(x)$ and $a(x)$ will increase proportional to
$a(x)$. The reaction processes will contribute to the equations with a
coupling term proportional to $a(x)n(x)$. It is also clear that monomer
flights will introduce terms proportional to $a(x)^2$, $a(x)^3$,
etc. Since only the lower order terms are relevant for a
renormalization group treatment \cite{cardyb}, we just keep the term
proportional to $a(x)^2$. Then, we can write down the following
Langevin equations: 
\begin{equation}
\label{lang}
\partial n(x,t) / \partial t = k_{1} \nabla^{2} a(x,t) + k_{2} a(x,t)
-k_{3} a(x,t)^{2} - k_{4} a(x,t) n(x,t) + \eta_{1} (x,t) 
\end{equation}

\begin{equation}
\label{lang1}
\partial a(x,t) / \partial t = u_{1} \nabla^{2} a(x,t) + u_{2} a(x,t)
-u_{3} a(x,t)^{2} - u_{4} a(x,t) n(x,t) + \eta_{2} (x,t) 
\end{equation}
where $\eta_{1} (x,t)$ and $\eta_{2} (x,t)$ are uncorrelated noises
proportional to $\sqrt{a(x,t)}$, $k_{i}$ and  $u_{i}$ are coefficients. 
This system of coupled Langevin equations is similar to that obtained
for the `pair contact process' \cite{Jens1,Jens2} which is one
of the prototype systems with multiple absorbing states.  Mu\~{n}oz
{\it et al} \cite{munioz1} have shown that for large $t$ the equation
corresponding to the activity (equation (\ref{lang1}) for the present model)
reduces to the Langevin representation of DP. Then, our simulation
results are consistent with the above-presented theoretical
arguments. The same authors have also shown that systems with many
available absorbing configurations display strong memory effects
that may lead to anomalous scaling. In addition to this, Mendes {\it et al} \cite{Mendes} have 
proposed a generalized hyperscaling relation which has proved to be valid in systems with 
multiple absorbing configurations.
Simulation results on several
lattice models with infinitely-many absorbing states support both
theoretical argument \cite{Mendes,Jens2,dick}. The role that initial states play in 
the temporal evolution of the present model is under investigation. 

\section{Conclusions and final remarks}
A model for the irreversible adsorption-reaction of a single species
on a continuous medium is studied by means of numerical
simulations. We would like to stress and comment upon the following
interesting features of the system: 
(i) in contrast to standard (reversible) transitions, non-equilibrium
IPT can happen in one dimension. (ii) the studied adsorption-reaction model
clearly shows interesting new effects that may arise when a
process is modelled on a continuous medium. Since the system always reaches a
stationary reactive state when simulated on a discrete lattice but a
final poisoned state can be observed on a continuous one (e. g. for
$P_r = 1$ and $R = 0$), one may expect a crossover behaviour when the
'discretization degree' of the surface is tuned from one extreme to
the other. This can be achieved by considering the adsorption on
\underline{discrete lattices} of species of arbitrary length $r$,
i. e. $r$-mers. We found that the reactive random sequential
adsorption (RRSA) of dimers ($r = 2$) always
leads to a reactive steady state, whose stationary coverage is
close to $\Theta \approx 0.5$, and no poisoning is observed. However,
the RRSA of trimers ($r = 3$) causes irreversible poisoning of the lattice with
an average saturation coverage close to $\Theta_{ST} \approx 0.7$. In
the case of dimers, two absorbing states of the type 
\begin{eqnarray}
...BBVBBVBBVBB... \\ \nonumber
...VBBVBBVBBVBBV... \nonumber
\end{eqnarray}
where $V$ is an empty site, can be expected. So, during the RRSA the formation
of several interfaces of the type $...BBVBBVVBBVBB...$ takes place. Due to
coarsening we expect that in the asymptotic limit ($t \rightarrow
\infty$) both absorbing states will form two semi-infinite domains
separated by an interface. The competition between these domains will
keep the system in the reactive state for ever. Just by increasing 
$r = 2$ to $r = 3$, an infinite number of absorbing configurations
appear in the system. So, the arguments developed above no longer hold
and poisoning is observed. Consequently, an IPT is located between $r = 2$ and $r
= 3$ and its precise location requires the study of the
adsorption-reaction problem with particles of non-integer
length. However, adsorption of $r$-mers of length $r = 2 + \epsilon$
would cause the argument of the two competing absorbing states to
fail. Therefore, we expect that the critical size, i. e. 'the
discretization degree' is $2$. (iii) To our best knowledge, this is the
first off-lattice model which exhibits a second-order IPT in the DP
universality class. Consequently, this results once again supports the
DP conjecture \cite{Janss} which can now be generalized to off-lattice
models with infinitely many absorbing states.   

We expect that the interesting behavior of the present simple reaction
model will stimulate further work on irreversible transitions and
critical phenomena on continuous media, a field that, to
our best knowledge, remains almost unexplored. 
\vskip 1.5 true cm
{\bf Acknowledgments}: This work was financially supported by CONICET,
UNLP, CIC (Provincia Bs. As.), ANPCyT, the Fundaci\'on Antorchas,
and the Volkswagen Foundation (Germany).

\newpage


\begin{figure}
\caption{a-c). Log-log plots of active area $A(t)$, survival
probability $P_s(t)$ 
and epidemic diameter $D(t)$ versus $t$. Results are obtained for a)
$P_r=0.60$ and $R=6.40$, b) $P_r = 0.80$ and $R=1.685$ and c) $P_r=1$ and
$R=0.853$. Averages are taken over ($10^5$) runs with different
poisoned initial states.}  
\label{fig1}
\end{figure}
\begin{figure}
\caption{Phase diagram for $B+B=0$ reaction on the continuum
(off-lattice). The active and adsorbing states zones are shown on a
$P_r$ versus $R/(R+1)$ 
plot. The extreme values of $R$ in the diagram are $0.853$ and
$\infty$ for $P_r=1$ 
and $P_r=0.5$, respectively. According to that shown in Figs. 1 and
3 the case  
given by the values $R= \infty$ and $P_r=0.5$ is the only one which does not
correspond to directed percolation universality class.}
\label{fig2}
\end{figure}
\begin{figure}
\caption{Log-log plot of the active area $A(t)$, survival
probability $P_s(t)$ and epidemic diameter $D(t)$ versus $t$ for the
singular case with $R= \infty$. The solution of this case
is exactly 
the one given by the mean field approximation. Dotted, full, and dashed lines
correspond to epidemic studies for $P_r <
0.5$, $P_r=0.5$ and  $P_r > 0.5$, respectively.}
\label{fig3}
\end{figure}




\begin{references}
\bibitem{Ligget} T. M. Ligget, {\it Interacting Particle Systems}
(Springer-Verlag New York, 1985).  
\bibitem{ZGB} R. Ziff, E. Gulari, Y. Barshad Phys. Rev. Lett. {\bf 56}, 2553,
(1986).
\bibitem{Eze1} E. V. Albano, Heter. Chem. Rev. {\bf 3}, 389, (1996).
\bibitem{Eze2} E. V. Albano and A. Maltz, Surf. Sci. {\bf 277}, 414,
(1992); E. V. Albano, J. Phys. A {\bf 25}, 2557, (1992).
\bibitem{Jens1} I. Jensen, Phys. Rev. Lett. {\bf 70}, 1465, (1993) and
J. Phys. A {\bf 27}, L61, (1994).
\bibitem{Jens2} I. Jensen and R. Dickman, Phys. Rev. E {\bf 48}, 1710, (1993);
Int. J. Mod. Phys. B {\bf 8}, 3299, (1994); R. Dickman, Phys. Rev. E
{\bf 53}, 2223, (1996).    
\bibitem{Eze3} E. V. Albano, J. Phys. A {\bf 27}, L881, (1994) and
Physica A {\bf 216}, 213, (1995).
\bibitem{Janss} H. K. Janssen,  Z. Phys. B {\bf 42}, 151, (1981); P. 
Grassberger, Z. Phys. B {\bf 47}, 365, (1982).
\bibitem{Grass1} P. Grassberger and A. de la Torre,  Ann. Phys. (New York) {\bf
122}, 373, (1979); P. Grassberger, J. Phys A: {\it Math Gen} {\bf 22}, 3673,
(1989).
\bibitem{Eze4} E. V. Albano, Physica A {\bf 214}, 424, (1995).
\bibitem{BARWO} H. Takayashu and A Y. Tretyakov, Phys. Rev. Lett. {\bf 68},
3060, (1992); I. Jensen, Phys. Rev. E {\bf 47}, 1, (1993).
\bibitem{Jens4} I. Jensen, H. Fogedby and R. Dickman Phys. Rev. A {\bf 41},
R3411, (1990).
\bibitem{Kolea} M. H. Kim and H. Park, Phys. Rev. Lett. {\bf 73}, 2579, (1994);
H. Park, M. H. Kim and H. Park, Phys. Rev. E {\bf 52}, 5664, (1995); 
H. Park and H. Park, Physica A {\bf 221}, 97, (1995).
\bibitem{Haye} H. Hinrichsen, Phys. Rev. E {\bf 55}, 219, (1997).
\bibitem{Roberto} R. A. Monetti, Phys. Rev. E {\bf 58}, 144, (1998).
\bibitem{Ponja} N. Inui, A Y. Tretyakov and H. Takayasu, J. Phys. A {\bf 28}, 1145, (1995).
\bibitem{Uwe} J. L. Cardy and U. C. T\"auber, Phys. Rev. Lett. {\bf
 77}, 4780, (1996); J. L. Cardy and U. C. T\"auber J. Stat. Phys. {\bf
90}, 1, (1998). 
\bibitem{Jim} J. W. Evans, Rev. Mod. Phys. {\bf 65}, 1281, (1993).
\bibitem{car} A. Renyi, {\it Sel. Trans. Math. Stat. Prob.} {\bf 4},
205, (1963).
\bibitem{Kim} J. Y. Kim and J. Lee, Phys. Rev. Lett. {\bf 82}, 1325, (1999).
\bibitem{priv} V. Privman and P. Nielaba, Europhys. Lett. {\bf 18}, (1992) 673.
\bibitem{eis} E. Eisenberg and A. Baram. J. Phys. A. {\bf30} (1997) L271.
\bibitem{stinc} M. Barma, M. D. Grynberg and R. B. Stinchcombe, Phys. Rev. Lett.
{\bf 70} (1993) 1033.
\bibitem{Daniel} D. H. Linares, R. H. Lopez and V. D. Pereyra,
J. Phys. A (Math. Gen.), {\bf 31}, 1165, (1998).
\bibitem{Mendes} J F F Mendes, R. Dickman, M. Henkel and M C. Marques, J. Phys. A  {\bf 27},
3019, (1994). 
\bibitem{series} I. Jensen, J. Phys. A {\bf 29}, 7013, (1996).
\bibitem{cardyb} J. Cardy, {\it Scaling and Renormalization in
Statistical Physics}, (Cambridge University Press, 1996).
\bibitem{munioz1} M. A. Mu\~noz, G. Grinstein, R. Dickman and R. Livi,
Phys. Rev. Lett. {\bf 76}, 451, (1996) and Physica D {\bf 103}, 485,
(1997).
\bibitem{dick} G. \'{O}dor, J. F. Mendes, M. A. Santos, and
M. C. Marquez, Phys. Rev. E {\bf 58}, 7020 (1998).
\end{references}
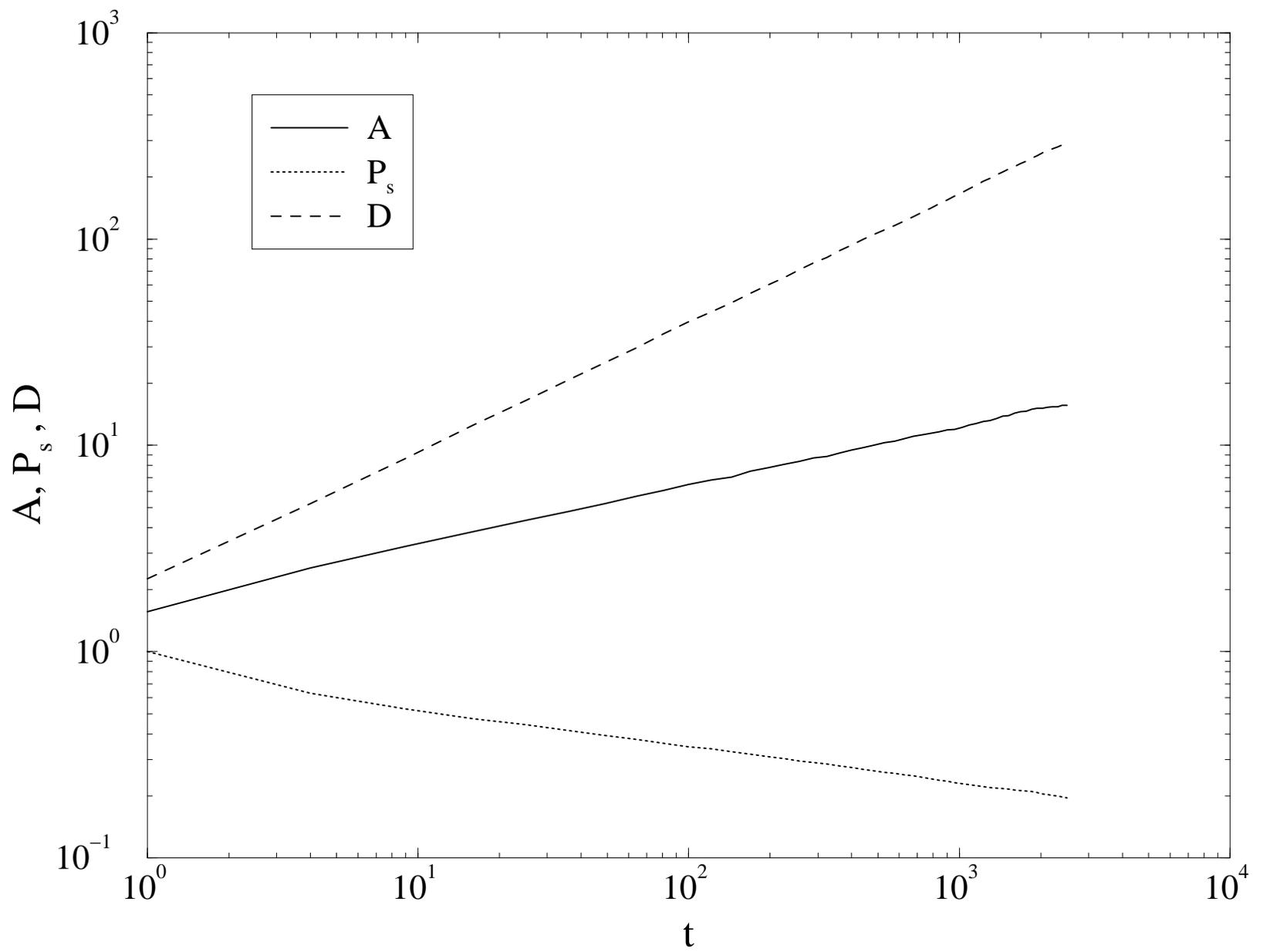
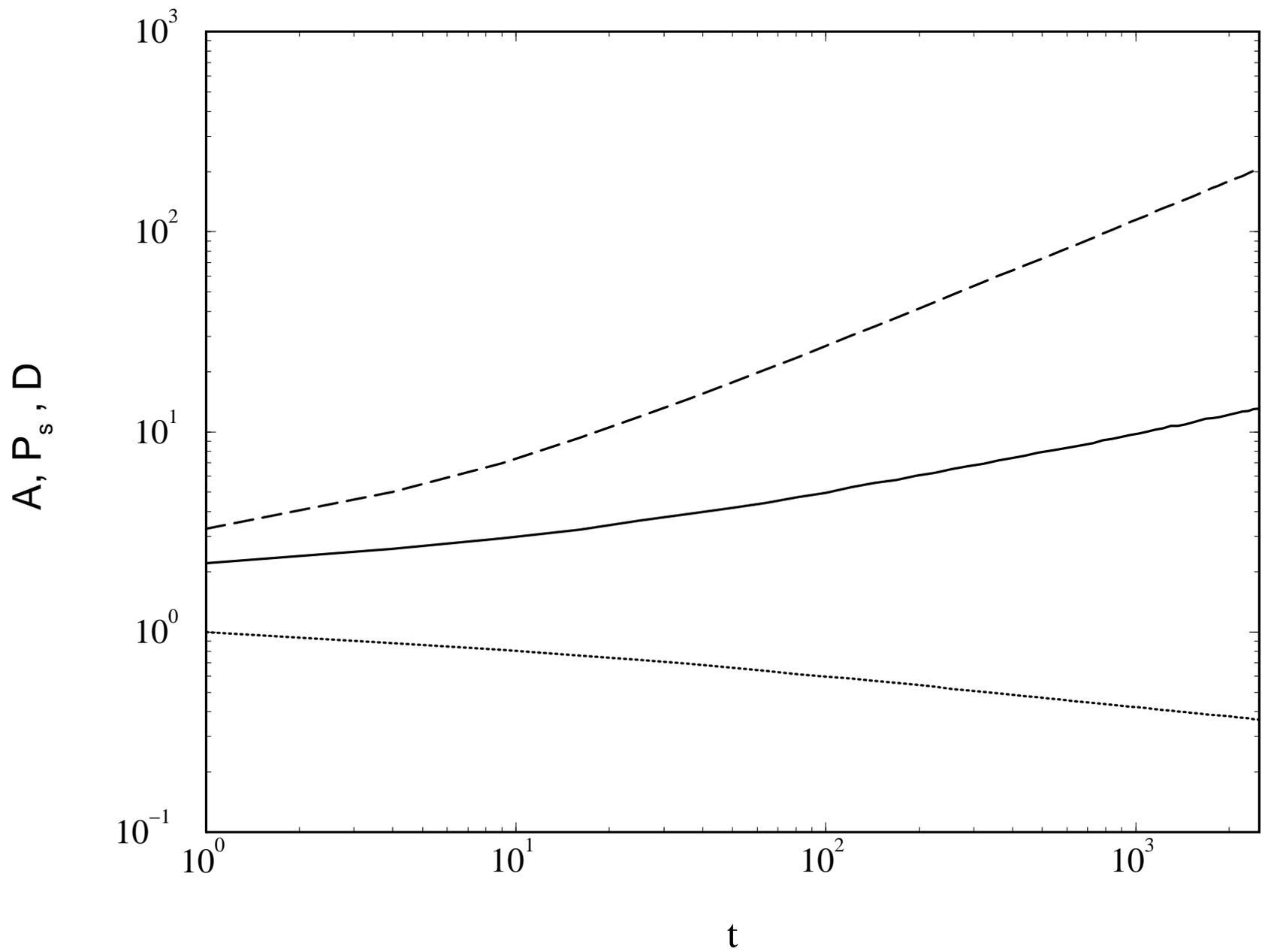
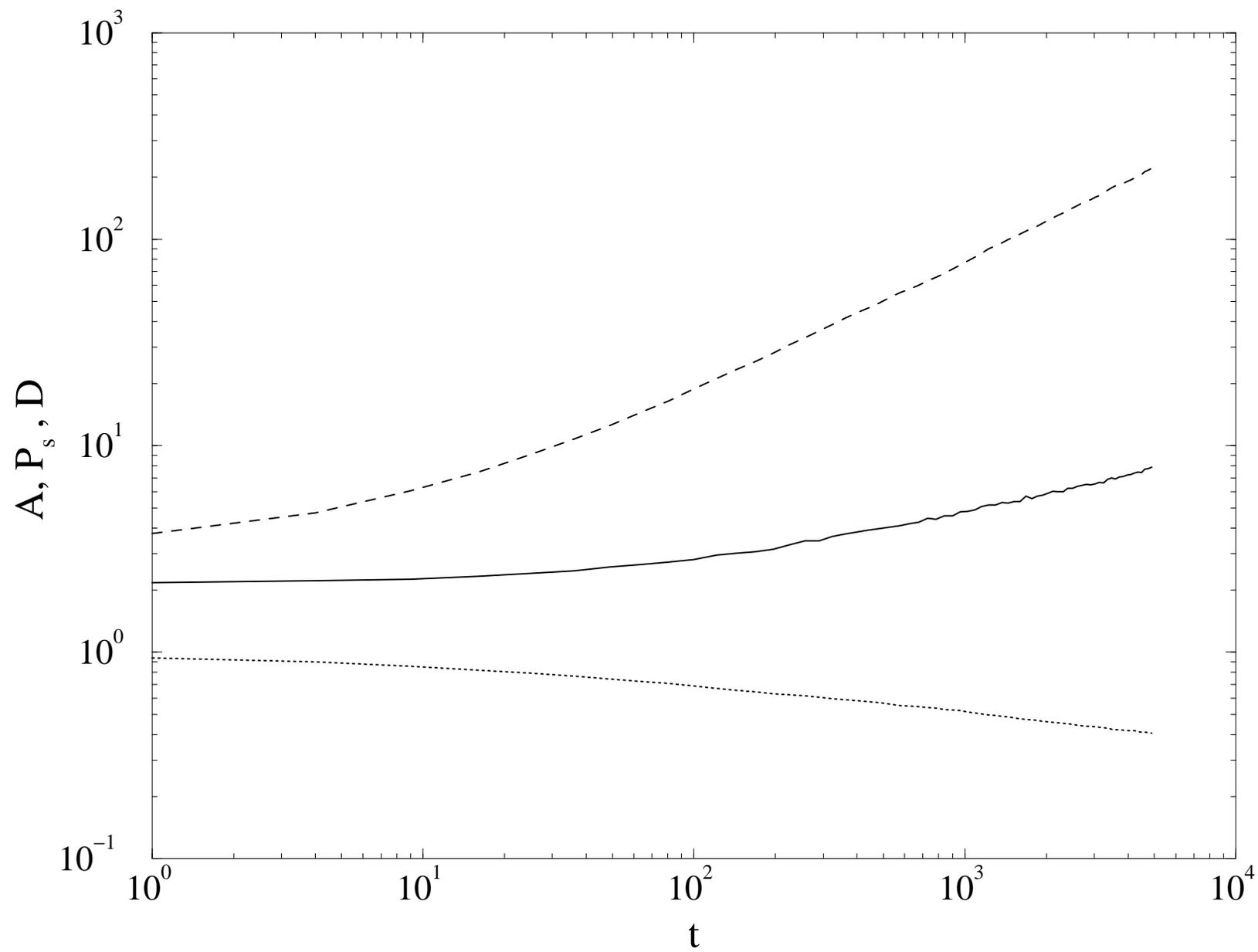
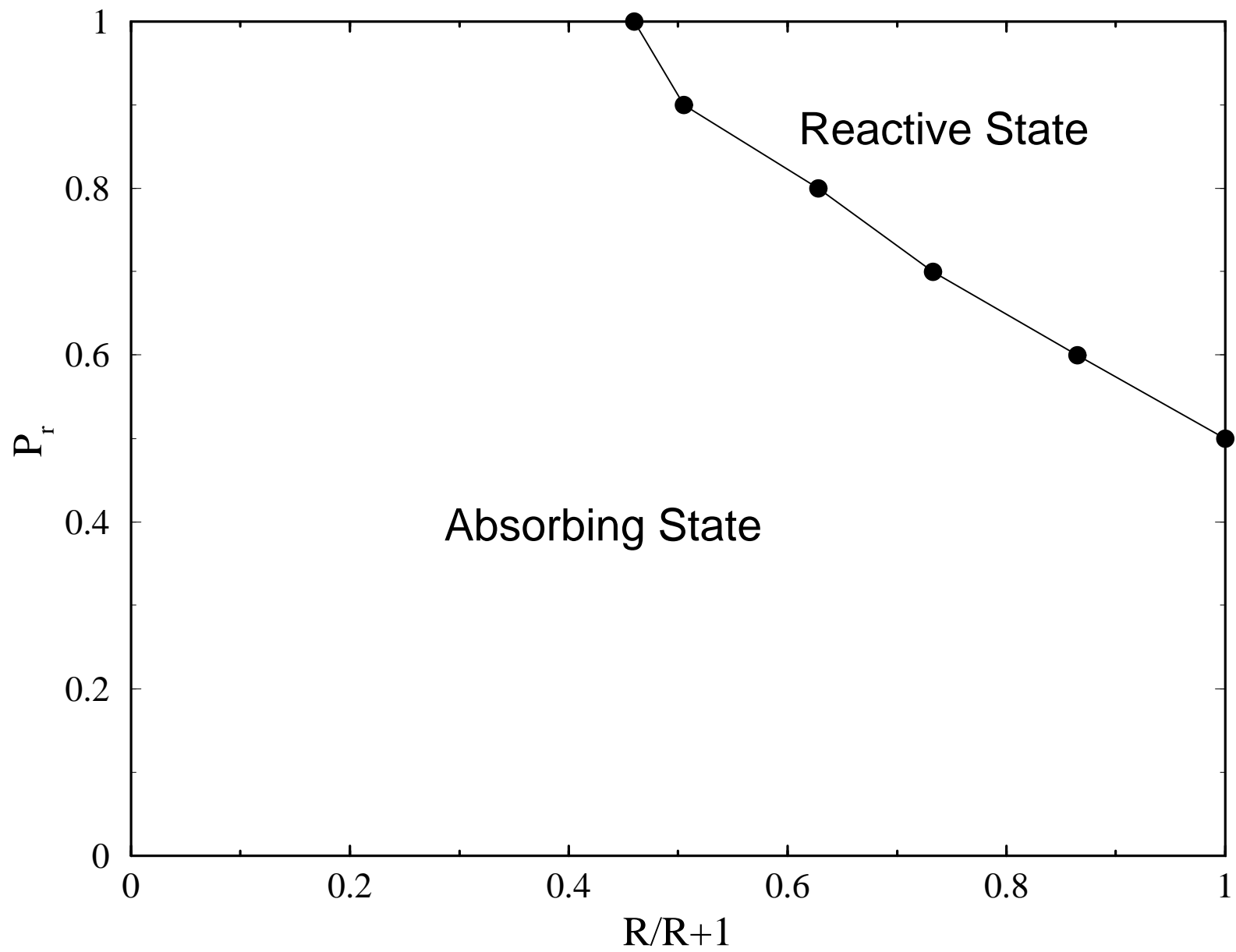
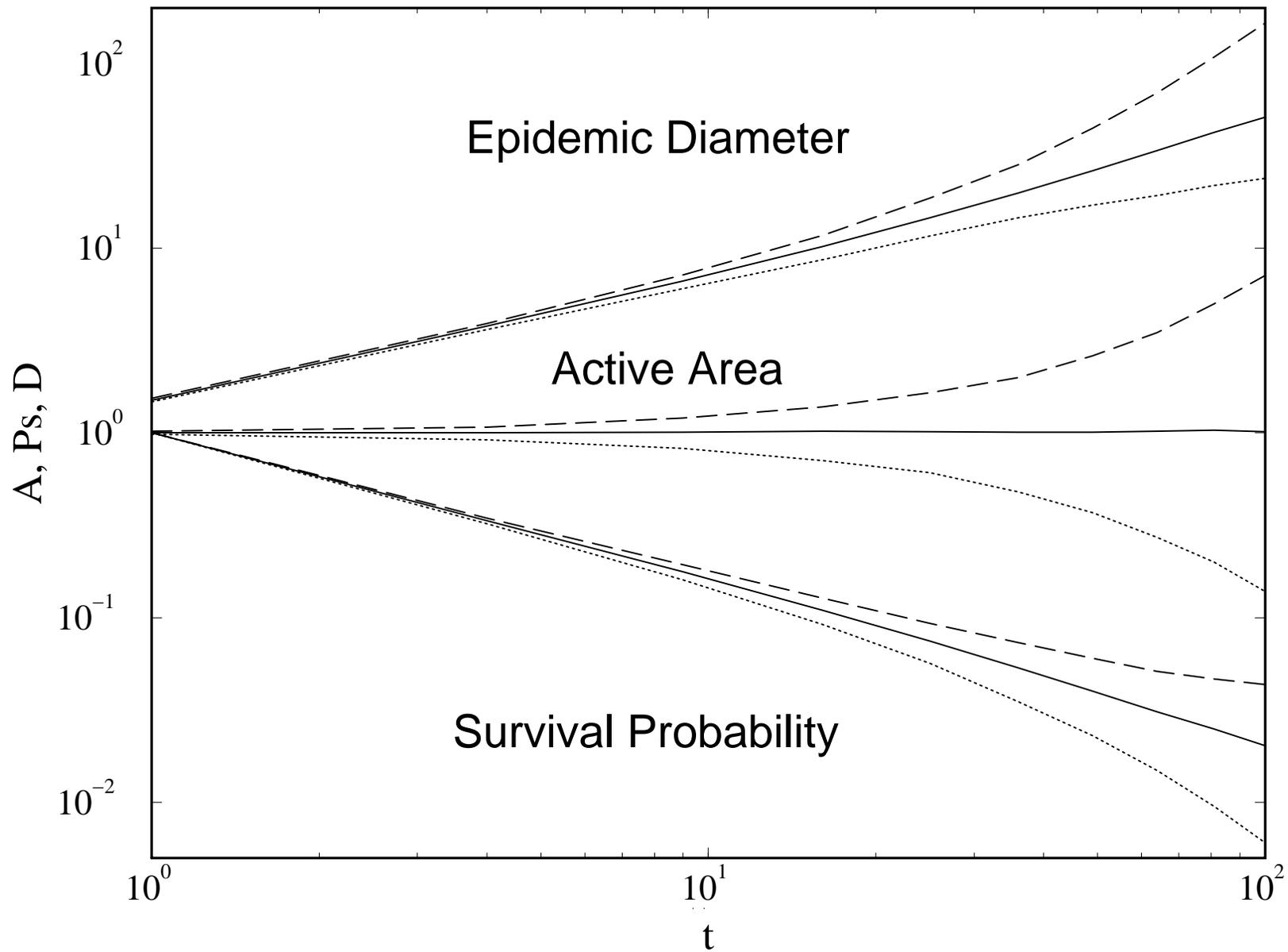
\end{document}